\documentstyle[12pt,linedraw]{article}
\newcommand{\tr}  {{\rm{tr}}}
\newcommand{\Trns}{{\scriptscriptstyle \rm T}}
\newcommand{\be}  {\begin{equation}}
\newcommand{\ee}  {\end{equation}}
\newcommand{\Cdots} {{\cdot\!\cdot\!\cdot\,}}

\begin{document}
\title{\bf A practical gauge invariant regularization
of the SO(10) grand unified model.}

\author{M.M.Deminov${}^1$ and A.A.Slavnov${}^{1,2}$.}

\maketitle

\centerline{\it ${}^1$Moscow State University, Moscow, 117234, Russia.}
\centerline{\it ${}^2$Steklov Mathematical Institute, Russian Academy of Sciences,}
\centerline{\it Gubkina 8, Moscow, 117966, Russia.}

\begin{abstract}
It is shown that a simple modification of the dimensional
regularization allows to compute in a consistent and gauge invariant way
any diagram with less than four loops in the SO(10) unified model.
The method applies also to the Standard Model generated by the
symmetry breaking SO(10) $\to$
${\rm SU(3)}\times{\rm SU(2)}\times{\rm U(1)}$.
A gauge invariant regularization for arbitrary diagram is also
described.
\end{abstract}

I. Introduction.

An invariant regularization is an important part of renormalization
of gauge invariant models. Although in principle any
regularization procedure may be used, symmetry breaking
regularization requires non gauge invariant counterterms, which
makes renormalization quite cumbersome.

In practical calculations the dimensional regularization
(see \cite{HV}) was mainly used so far.
However, the dimensional regularization is not applicable to the
models with chiral fermions as in this method there is no
symmetry preserving consistent definition of $\gamma_5$-matrix.
This means in particular that the dimensional regularization
does not preserve the symmetry of the Standard Model and grand
unified models which reduce at low energies to the Standard Model.
At one loop level it is not a very serious problem, as one can find
relatively easy counterterms which restore gauge invariance.
However for diagrams including higher loops this procedure
becomes more and more complicated.

Another gauge invariant regularization is provided by the higher
covariant derivative (HCD) method \cite{Sl1, LZ}.
This regularization has the advantage of being implementable at
the Lagrangian level practically to any model. However it does not
provide a complete regularization: one loop diagrams remain
divergent. For pure Yang-Mills theory the additional gauge
invariant regularization of one loop diagrams was proposed in
reference \cite{Sl2}. Some problems related to this additional
regularization were discussed in references \cite{MR, AF}.
The complete and self-consistent procedure of HCD regularization
is described in the paper \cite{BSl} (see also \cite{B}).

However for models with chiral fermions the problem still existed
as no invariant regularization for the one loop fermion diagrams
was known. For the case of the Standard Model this problem was
solved in the paper \cite{FSl}. The procedure developed in \cite{FSl}
combined with the HCD method of the paper \cite{BSl} provides a
complete gauge invariant regularization of the Standard Model.

Nevertheless for practical calculations this procedure is not very
convinient due to a complicated structure of regularized
Lagrangian and a simpler method would be welcome.

Theoretical analysis of the present experiments on searches of
Higgs meson and determination of quark mass requires calculation
of two loop diagrams, so a construction of a simple symmetry
preserving regularization is of great practical importance.

In the present paper we propose a new gauge invariant
regularization method for the unified SO(10) model which combines ideas
of different approaches \cite{HV, Sl1, FSl}.
For the one, two and three loop diagrams which at present are the most
important from the practical point of view we demonstrate that
a simple modification of dimensional regularization
is sufficient to provide a gauge invariant calculation procedure.

Our method is also applicable to the Standard Model as it may be
obtained from the unified SO(10) model via breaking SO(10) gauge group
to ${\rm SU(3)}\times{\rm SU(2)}\times{\rm U(1)}$.
This symmetry breaking may occur spontaneously via Higgs mechanism
if one insists on the unification of all interactions at very
high energies, or may be introduced explicitly if one is
interested only in study of the Standard Model at low and intermediate
energies. In this paper we show that such symmetry breaking is compatible
with our method. A detailed discussion will be given elsewhere.

For a general multiloop diagram this method is not sufficient
(see for example diagrams on figures 4 and 5 in Appendix).
In this case the gauge invariant regularization may be achieved by
combining the dimensional regularization with the HCD method.
It is important to notice that in our case it is sufficient simply
to introduce higher covariant derivatives in the Lagrangian and no
additional problems with one loop diagrams arise. In this way one
can avoid the most complicated part of the HCD method and
calculations remain reasonably simple.

\medskip II.
Modified dimensional regularization for the diagrams
with less then four loops.

The difficulties of applying dimensional regularization to
chiral models are due to impossibility of a consistent symmetry
preserving definition of the $\gamma_5$-matrix. The usual
definition
$\gamma_5=\frac{i}{4!}\varepsilon^{\mu\nu\rho\sigma}\gamma_\mu\gamma_\nu\gamma_\rho\gamma_\sigma$
involves the totally antisymmetric tensor
$\varepsilon^{\mu\nu\rho\sigma}$,
which has no natural extension beyond $d=4$.
One can try to define $\gamma_5$ in arbitrary dimension
axiomatically, postulating that it anticommutes with all
$\gamma_\mu$-matrices (see e.g. \cite{Ch}). It was shown however that such a definition
contradicts the cyclicity property of $\gamma$-matrices trace
(see e.g. \cite{Col}).

Another definition was proposed by G.'tHooft and M.Veltman \cite{HV}
who postulated that $\bar \gamma_5=i\gamma_0\gamma_1\gamma_2\gamma_3$
in arbitrary dimension. This definition is obviously
self-consistent, but due to the fact that $\bar \gamma_5$
anticommutes with $\gamma_\mu,\; \mu=0,1,2,3,$ and commutes with
all other matrices, it breaks chiral invariance and hence the
gauge invariance.
Of course, the gauge
invariance of renormalized theory may be restored by making
subtractions in such a way to provide Slavnov-Taylor identities
\cite{Sl3,T} for renormalized Green functions, but it complicates
the job drastically. The detailed analysis of this approach was
carried out by P.Breitenlohner and D.Maison. \cite{BM}.
The symmetry breaking in this approach arises from two sources.
Any convergent diagram $\Pi_c$ may acquire a noninvariant piece,
vanishing at $d=4$:
\be
\Pi_c=\Pi^{inv}_c+(d-4)\tilde\Pi.
\label{Sm1}
\ee
The term $\tilde \Pi$ by itself is harmless as long as the
diagram $\Pi$ is not a subgraph of some divergent diagram. In
the later case $\Pi_c$ is multiplied by the pole term proportional
to $1/(d-4)$ and $\tilde\Pi$ gives a nonzero symmetry breaking
contribution.

For divergent diagram $\Pi_d$ the limit $d\to4$ does not exist and
the diagram by itself may acquire a finite or infinite
noninvariant term
\be
\Pi_d=\Pi_p^{inv} + \frac{1}{d-4}\Pi'+\Pi''
\label{Sm2}
\ee
We stress that with this prescription breaking of gauge invariance
may occur not only in fermion loop, but also in diagrams involving
open fermion lines.

On the contrary, the prescription of total anticommutativity meets
no problems in diagrams without fermion loops and preserves gauge
invariance. In this case the problem is related solely to fermion
loops.

Below we shall show that for any diagram with less then four loops
in the SO(10) unified model as well as for a large class of higher
loop diagrams this problem does not arise as these diagrams in fact
do not depend on $\gamma_5$ matrix properties in dimensions $d\ne 4$.
The proof will essentially follow the ideas of the paper \cite{FSl}.

The gauge invariant $\rm{SO(10)}$ Lagrangian
with spinor fields is chosen in the form:
\be
{\cal L} =
-\frac{1}{4}F_{\mu\nu}^2 +
i \overline\psi\gamma_\mu (\partial_\mu - i {\sl g} A^{ij}_\mu \sigma_{ij})\psi.
\label{SO10Main}
\ee
Here the spinors $\psi$ span the 16 dimensional representation of
$\rm{SO(10)}$.
The spinor fields have positive chirality with respect to the Lorentz group:
$\psi = \frac{1}{2}(1+\gamma_5)\psi.$
The spinors $\psi$ include left components of quarks
and leptons and their charge conjugated right components.
The fifteen components of $\psi$ describe exactly one
generation of the Standard Model. The remaining component
corresponds to the right-handed neutrino. It is a singlet
with respect to ${\rm SU(3)}\times{\rm SU(2)}\times{\rm U(1)}$
group, and if one is interested only in the description of the Standard
Model, this component may be omitted.

The 16-dimensional irreducible representation of SO(10) is
obtained by applying the projector to $32$-dimensional vector:
$\psi=\frac{1}{2}(1+\Gamma_{11})\psi.$
The matrices $\sigma_{ij}$ are the $\rm{SO(10)}$ generators:
$\sigma_{ij} = \frac{i}{2}[\Gamma_i,\Gamma_j],$ where $\Gamma_i$ are
Hermitian $32$ by $32$ matrices which satisfy the Clifford algebra:
$\{\Gamma_i,\Gamma_j\} = 2\delta_{ij};$
$\Gamma_{11} = {-i \Gamma_1 \Gamma_2 \cdots \Gamma_{10}}.$
Structure of these $32$ by $32$ matrices is presented in \cite{Matr}.

The ${\rm SU(3)}\times{\rm SU(2)}\times{\rm U(1)}$ invariant
Lagrangian may be obtained either by putting all gauge fields
except for the ones corresponding to the generators of this
subgroup equal to zero, or by introducing proper Higgs fields.
Further reduction of the symmetry group to ${\rm SU(3)}\times{\rm U(1)}$
also may be provided by condensate of Higgs fields. We postpone a
detailed discussion of this procedure till special publication,
but demonstrate now that introduction of Higgs fields interaction
may be easily incorporated in our method. In particular the masses
for W and Z mesons are generated by the following Higgs field
Lagrangian:
\be
{\cal L}_H=\frac{1}{32}\tr (D_\mu\phi)^\dag D_\mu\phi
-\frac{1}{2}\psi^\Trns C_D C (\phi+\phi^\dag) \psi
+\frac{1}{2}\bar\psi C_D (\phi+\phi^\dag) C \bar\psi^\Trns
-\lambda(\tr(\phi^\dagger\phi)-\mu^2)^2.
\label{LHiggs}
\ee
Here the Higgs field $\phi$ is $32\times 32$ matrix:
$\phi=\phi_i\Gamma_i,\, \phi_i$ are complex, $i=1\ldots10$ (here and below summation over
repeated indices is assumed),
$\;D_\mu \phi = \partial_\mu\phi-i{\sl g}[A_\mu^{ij}\sigma_{ij},\phi].$
The matrix $C_D$ is a charge conjugation matrix.
The matrix $C$ is a conjugation matrix defined by the relation
$\sigma^{\Trns}_{ij} C = - C \sigma_{ij}$, which
provides gauge invariance of second and third terms
in (\ref{LHiggs}). Matrix $C$ anticommutes with
$\Gamma_{11}:\,$ $C\Gamma_{11}=-\Gamma_{11}C.$

Spontaneous symmetry breaking is due to nonzero vacuum expectation
value of the Higgs field $\langle \phi \rangle$:
\be
\langle \phi \rangle =\frac{\mu}{4\sqrt{2}}(\alpha \Gamma_3+\beta\Gamma_4),
\quad \alpha\alpha^\dag+\beta\beta^\dag=1.
\label{Sm7}
\ee

The usual perturbation theory arises after the shift of Higgs
fields by the stationary value (\ref{Sm7}). However for a general
analysis it is more convinient to work in terms of unshifted
fields. Then the mass terms arise due to interaction with
condensate: summation of the series of insertions of arbitrary
number of scalar vertices describing the interaction
of fermion, scalar or vector fields with the condensate results in the shift
of the masses in the corresponding propagators.

The useful observation is that the theory described by the
Lagrangian (\ref{SO10Main}), (\ref{LHiggs}) is invariant
under simultaneous change of signs of $\gamma_5$ and $\Gamma_{11}$
matrices. Indeed, these Lagrangians may be rewritten in terms of
transposed fields:
\be
{\cal L}_\psi=-i\partial_\mu\psi^\Trns \frac{(1+\gamma_5^\Trns)}{2}
\frac{(1+\Gamma_{11}^\Trns)}{2}\gamma_\mu^\Trns\bar\psi^\Trns-
{\sl g}\psi^\Trns\frac{(1+\gamma_5^\Trns)}{2}
\frac{(1+\Gamma_{11}^\Trns)}{2}
\sigma_{ij}^\Trns A_\mu^{ij} \gamma_\mu^\Trns
\bar\psi^\Trns.
\label{Sm8}
\ee
Let us multiply $\psi^\Trns$ by the unit factor $CC_DCC_D$.
Commuting $CC_D$ to the right we get
$$
{\cal L}_\psi=i\psi^\Trns CC_D\frac{(1+\gamma_5)}{2}
\frac{(1-\Gamma_{11})}{2}(-\gamma_\mu)CC_D\partial_\mu\bar\psi^\Trns-
$$
\be
-{\sl g}\psi^\Trns CC_D\frac{(1+\gamma_5)}{2}
\frac{(1-\Gamma_{11})}{2}
(-\sigma_{ij}) A_\mu^{ij} (-\gamma_\mu)
CC_D\bar\psi^\Trns.
\label{Sm9}
\ee
Introducing the conjugated fields $\psi^c=CC_D\bar\psi^\Trns$,
one can write this equation in the form:
\be
{\cal L}_\psi=
-i\bar\psi^c\gamma_\mu(\partial_\mu-i{\sl g}A_\mu^{ij}\sigma_{ij})
\frac{(1-\gamma_5)}{2}\frac{(1-\Gamma_{11})}{2}\psi^c.
\label{Sm10}
\ee
Similarly, the Higgs-fermion interaction Lagrangian is rewritten
as
\be
{\cal L}_{\phi\psi}=
-\frac{1}{2}{\psi^c}^{{}^\Trns} C_D C (\phi+\phi^\dag)
        \frac{(1-\gamma_5)}{2}\frac{(1-\Gamma_{11})}{2}\psi^c
+\frac{1}{2}\bar\psi^c C_D (\phi+\phi^\dag) C
        \frac{(1+\gamma_5)}{2}\frac{(1-\Gamma_{11})}{2}\bar{\psi^c}^\Trns,
\label{Sm11}
\ee
It follows from equations (\ref{Sm10}),(\ref{Sm11}) that
simultaneous change of the signs of $\gamma_5$ and $\Gamma_{11}$
does not influence the value of any diagram. In particular, if a
diagram does not involve $\Gamma_{11},$ it may be written in the
form which also does not involve $\gamma_5$-matrix.
This property is very important for us,
and it will be used below.

Now let us consider arbitrary one loop fermion diagram in the
model described by Lagrangian (\ref{SO10Main}),(\ref{LHiggs}).
Any such loop is proportional to
\be
\tr\left[\frac{(1+\Gamma_{11})}{2}\sigma^{i_1i_2}\cdots\Gamma^{i_k}\cdots
 \sigma^{i_{n-1}i_n}\right]
\label{Sm12}
\ee
-- a trace of a product of the projector $\frac{1}{2}(1+\Gamma_{11})$
and $\sigma$- and $\Gamma$-matrices, which correspond to external
vector and scalar lines respectively.
Each $\sigma$ matrix is a product of two different $\Gamma$
matrices. Hence the trace of the term, proportional to $\Gamma_{11}$
is zero if the number $n$ is less than 10,
in other words the $\Gamma_{11}$ under the trace is multiplied
by less then 10 $\Gamma$-matrices.
That means that all one loop fermion diagrams for which the sum of
the number of vector external lines and one half of the number
of scalar external lines is less then 5 do not involve $\Gamma_{11}$
and therefore do not depend
on $\gamma_5.$

Moreover one can easily see that these diagrams up to the factor
$1/2$ coincide with the corresponding diagrams in the model where
the fermions span the $32$-dimensional representation of SO(10).
This model is vectorlike and may be rewritten in a purely
vectorial form.

It is proven in the Appendix that this property
(absence of $\gamma_5$) remains valid for
a fermion loop with arbitrary boson insertions
if the sum of the number of vector external lines
and one half of the number of scalar external lines
is less then 5.

That means one can apply to such diagrams dimensional
regularization with the prescription of total anticommutativity
of $\gamma_5$ matrix. The procedure is manifestly gauge invariant,
and no problems with the $\gamma$-matrices traces arise.

A slight modification of this discussion allows to extend our
procedure to the diagrams with arbitrary number of external vector
or scalar lines. All diagrams with one fermion loop for which
the sum of the number of vector external lines and one half of the number of
scalar external lines is more then 5, that is the diagrams with
10 or more SO(10)-group indices corresponding to external vector
or scalar lines are superficially convergent.
To renormalize such a diagram it is
sufficient according to R-operation to subtract the counterterms
corresponding to all divergent subgraphs. As was discussed above
these subgraphs may be calculated using the dimensional regularization
and minimal subtractions.
After such subtraction the  corresponding integral
becomes convergent and one can remove the dimensional
regularization before calculating the trace over the fermion loop,
where $\gamma_5$ can give nonzero contribution. The calculation of
this trace has to be carried out in four-dimensional space-time,
hence there are no problems with $\gamma_5$.

This remark is particularly important for taking into account the
Higgs field condensate contribution. We remind that in our
discussion we assume that the fermion masses in propagators arise
via summation of insertions of Higgs field condensate to fermion
lines. That means some ''external'' Higgs field line are fictitious
and amount only to the shift of fermion masses. These insertions
are proportional to $\alpha \Gamma_3+\beta\Gamma_4$ and several
insertions may generate additional factors $\sigma_{34}$ in the
trace (\ref{Sm12}). Presence of such factors do not change our
reasoning for diagrams with two or three ''real'' external lines.
To make the traces in these diagrams different from zero,
one needs at least two different $\sigma$-matrices. The mass
insertion cannot provide two different $\sigma$-matrices.

The situation is different for the diagrams with four ''real''
external lines. In this case additional matrix $\sigma_{ab}$
arising from mass insertions can make the trace with factor $\gamma_5$
different from zero. However such a diagram is superficially
convergent and as was discussed above also allows the dimensional
regularization. Obviously analogous arguments may be applied if
one works in terms of ''shifted'' Lagrangian generating massive
propagators.

The same arguments are applicable to other Higgs field
interaction: insertion of Higgs field condensates to fermion lines
either do not introduce $\gamma_5$ matrix, or makes the diagram
finite.

Another  class of diagrams which allow a straight-forward
dimensional regularization with the $\gamma_5$ matrix defined via
anticommutativity with all $\gamma_\mu$ consists of the diagrams without
fermion loops. In this case there is no need to calculate the
trace over spinoral indices and the gauge invariance is preserved.

So we showed that a large class of the diagrams
in the unified SO(10) model may be computed using
the dimensional regularization
with additional prescriptions formulated above.

The diagrams which do not fall into this class are presented by
the superficially divergent diagrams where open fermion line
or fermion loop for which the sum of number of vector
lines and the one half of the number of scalar lines is more then 5
is linked to a
fermion loop for which the sum of number of vector
lines and the half of the number of scalar lines is more then 5.
(see figures 4 and 5 respectively).
The lowest order diagrams of this type include 4 loops.

To deal in a gauge invariant way with these diagrams one can use
a hybrid regularization which combines the dimensional
regularization and HCD method.

\medskip
III. The Hybrid gauge invariant regularization.

There are several possibilities for combining the dimensional
regularization with HCD method. A particular choice may be done on
the basis of the most simple calculations of the diagram in
question. As at the moment calculations of four and higher loop
diagrams in the Standard Model are of mainly academic interest we
present here only one, conceptually the most straight-forward
method.

The regularized Lagrangian may be chosen in the form:
$${\cal L}_\Lambda=
-\frac{1}{4}(F_{\mu\nu}^2+
\frac{1}{\Lambda^2}D_\alpha F_{\mu\nu}D_\alpha F_{\mu\nu})+
i\bar\psi\gamma_\mu D_\mu\psi
+\frac{1}{2}{\tr} (D_\mu\phi)^\dag(D_\mu\phi)+
$$
\be
+\frac{1}{\Lambda^2} (D^2\phi)^\dag(D^2\phi)
-\frac{1}{2}\psi^\Trns C_D C (\phi+\phi^\dag) \psi
+\frac{1}{2}\bar\psi C_D (\phi+\phi^\dag) C \bar\psi^\Trns
-\lambda({\tr}(\phi^\dag\phi)-\mu^2)^2.
\label{LHCD}
\ee

In the model described by the Lagrangian (\ref{LHCD}) all multiloop
diagrams are superficially convergent. One loop diagrams with
external fermion lines are convergent too. So the only diagrams
which need additional regularization are the one loop diagrams
without external fermion lines, in particular fermion loops with
less then five external fields. As was discussed above these
diagrams do not depend on $\gamma_5$ and may be described by the
effective vectorlike model. Calculation of an arbitrary diagram
proceeds as follows. Keeping $\Lambda$ finite one subtracts
the counterterms corresponding to divergent subgraphs using
dimensional regularization and the minimal subtractions.
As these subdiagrams do not involve $\gamma_5$ no problem arises.
After that the diagram becomes finite at the dimension
$d=4$. (We remind that the $\Lambda$ is still finite.)

The second step is removing the higher derivative regularization
by taking the limit $\Lambda\to\infty$. This procedure is also
manifestly gauge invariant and no symmetry breaking counterterms
are needed.

Introduction of higher derivatives makes the calculations in this
method more complicated. However using the dimensional
regularization allows to avoid the additional regularization
of one loop diagrams which is the most cumbersome part of the HCD
method.
\par\medskip IV. Discussion.

The procedure described in the section II provides a simple
practical method of gauge invariant computation of all
diagrams with less then four loops in the unified SO(10) model.
This method also applies to the Standard Model which may be
obtained via spontaneous symmetry breaking
${\rm SO(10)}\to{\rm SU(3)}\times{\rm SU(2)}\times{\rm U(1)}
\to {\rm SU(3)}\times{\rm U(1)}$.
Whether it is possible to modify further dimensional regularization
to make it applicable to an arbitrary diagram is at present the
open question. This question is under investigation.
Meanwhile arbitrary diagrams may be treated in a gauge invariant
way by means of the hybrid regularization described
in section III. We conclude by noticing that the gauge invariance
of our procedure was checked at the one loop level by explicit
calculation of gluon-W scattering in the Standard Model
\cite{DSl}.
\pagebreak

Acknowledgements.

One of the authors (A.A.S) is grateful to P.Breitenlohner and
D.Maison for helpful discussions. M.M.D is grateful
to M.M\"uller-Preussker for warm hospitality in the
Humboldt University of Berlin and to F.Jegerlehner for useful remarks.
This work is supported in part by the RBRF grant 99-01-00190,
grant for support of leading scientific schools 00-15-96046
and by Leonhard Euler scholarship grant.

\section*{Appendix}
Theorem :
If a diagram contains one fermion loop and no open fermion lines
(see figure~1)
then matrix $\Gamma_{11}$ gives a nonzero contribution only if
the sum of the number of external vector lines and one half of the number
of external scalar lines is equal to 5 or more.

\medskip
\scriptsize
\Lengthunit=1.0cm
\GRAPH(hsize=9){%
\Linewidth{0.5pt}%
\mov(0.2,3.6){\wavelin(1.8,0)}%
\mov(2.02,3.62){\vdot}%
\mov(1.6,3.8){$b_1b_2$}%
\mov(0.2,1.33){\wavelin(1.8,0)}%
\mov(2.04,1.34){\vdot}%
\mov(1.6,1){$b_3b_4$}%
\mov(4,1.33){\wavelin(1.8,0)}%
\mov(3.96,1.34){\vdot}%
\mov(4,1){$b_5b_6$}%
\mov(4,3.6){\wavelin(1.8,0)}%
\mov(3.98,3.62){\vdot}%
\mov(4,3.8){$b_{2n-1}b_{2n}$}%
\Linewidth{0.8pt}%
\mov(3,2.5){\Circle(3)}%
\mov(3,2.5){\Circle*(1)}%
\Linewidth{0.5pt}%
\mov(1.7,3.2){\wavelin1.3(0.8,-0.2)}%
\mov(1.9,3.2){$i_1i_2$}%
\mov(1.68,3.23){\vdot}%
\mov(4.32,3.23){\vdot}%
\mov(1.68,1.77){\vdot}%
\mov(4.32,1.77){\vdot}%
\mov(1.7,1.8){\wavelin1.3(0.8,0.2)}%
\mov(1.6,2.1){$i_4i_5$}%
\mov(4.3,1.8){\wavelin1.3(-0.8,0.2)}%
\mov(3.5,2.1){$i_7i_8$}%
\mov(4.3,3.2){\wavelin1.3(-0.8,-0.2)}%
\mov(3.5,2.8){$i_9i_{10}$}%
\mov(3,4){\wavelin1.3(0,-0.77)}%
\mov(3,4){\vdot}%
\mov(3,1){\dashlin1.3(0,+0.75)}%
\mov(3,1){\vdot}%
\mov(3.08,1.5){$i_6$}%
\mov(1.5,2.5){\dashlin1.3(0.75,0)}%
\mov(1.1,2.4){$i_3$}%
\mov(0.2,2.9){\dashlin1.3(1.0,0)}%
\mov(1.57,2.915){\vdot}%
\mov(1.1,3){$m_1$}%
\mov(4.45,2.1){\dashlin1.3(1.0,0)}%
\mov(4.45,2.12){\vdot}%
\mov(4.6,2.2){$m_s$}%
\mov(3.5,2.6){\wavelin(2.3,0)}%
\mov(4.7,2.75){$b_7b_8$}%
\mov(2.4,4.1){$i_{2s-1}i_{2s}$}%
\mov(2.76,2.3){\Large G}%
\mov(2.2,0.3){\normalsize Figure 1.}%
\Linewidth{0.5pt}%
\mov(7.2,3.6){\wavelin(1.8,0)}%
\mov(9.02,3.62){\vdot}%
\mov(8.3,3.8){$b_1b_2$}%
\mov(7.2,1.33){\wavelin(1.8,0)}%
\mov(9.04,1.34){\vdot}%
\mov(8.6,0.9){$b_rb_{r+1}$}%
\mov(7.2,2.7){\dashlin1.2(1.04,0)}%
\mov(8.492,2.7){\vdot}%
\mov(12.8,2.5){\dashlin1.2(-1.04,0)}%
\mov(11.5,2.5){\vdot}%
\mov(11.6,2.6){$m_r$}%
\mov(11,1.33){\wavelin(1.8,0)}%
\mov(10.96,1.34){\vdot}%
\mov(7.9,2.85){$m_1$}%
\mov(11,0.9){$b_sb_{s+1}$}%
\mov(11,3.6){\wavelin(1.8,0)}%
\mov(10.98,3.62){\vdot}%
\mov(11,3.8){$b_{2n-1}b_{2n}$}%
\Linewidth{0.8pt}%
\mov(10,2.5){\Circle(3)}%
\Linewidth{0.5pt}%
\mov(10,4){\wavelin(0,-3)}%
\mov(10,4){\vdot}%
\mov(10,1){\vdot}%
\mov(8.6,3.0){\wavelin1.32(2,-1)}%
\mov(8.6,3.0){\vdot}%
\mov(11.24,1.68){\vdot}%
\mov(10.5,3.9){\wavelin0.93(1,-2)}%
\mov(10.5,3.9){\vdot}%
\mov(11.43,2.04){\vdot}%
\mov(8.6,2.0){\dashlin1.32(2,1)}%
\mov(8.6,2.0){\vdot}%
\mov(11.24,3.32){\vdot}%
\mov(9.8,4.2){$ij$}%
\mov(9.5,1.3){$ij$}%
\mov(9.3,0.3){\normalsize Figure 2.}%
}
\normalsize \rm
Here and in the following dash lines correspond to Higgs particles.

The proof is based on the following identity:
\be
\sum_i
\Gamma_i\Gamma_{b_1}\Cdots\Gamma_{b_t}\Gamma_i=
2\sum_{l=1}^{t}(-1)^{(l+1)}
\Gamma_{b_1}\Cdots\Gamma_{b_{l-1}}
\Gamma_{b_{l+1}}\Cdots\Gamma_{b_t}\Gamma_{b_l}
+10 (-1)^l \Gamma_{b_1}\Cdots\Gamma_{b_t}.
\label{simp0}
\ee
It is important to note that the number of $\Gamma$'s in each
term of the r.h.s. is equal to the number of $\Gamma$'s in the original product
$\Gamma_{b_1}\cdots\Gamma_{b_t}$
and the indices of $\Gamma$'s belong to the same set.

Obviously an analogous representation holds if we multiply a
product of $\sigma_{b_1b_2}..\sigma_{b_{2t-1}b_{2t}}$ by
$\Gamma_i$ from both sides and sum over $i$.
Indeed,
\be
\Gamma_{b_i}\sigma_{b_ib_j}\Gamma_{b_i}=\sigma_{b_jb_i},\;\;
\mbox{and}\;\Gamma_{b_k}\sigma_{b_ib_j}\Gamma_{b_k}=\sigma_{b_ib_j}
\;\mbox{if $b_k\ne b_i,b_j$.}
\label{Sla1415}
\ee
The product
\be
\sigma_{b_1b_2}\cdots\Gamma_{m_1}\cdots
\sigma_{b_lb_{l+1}}\cdots\Gamma_{m_s}\cdots
\sigma_{b_{2t-1}b_{2t}}
\label{Sla16}
\ee
after multiplication by $\Gamma_i$ from both sides
and summation over $i$ may be represented as a sum of products
$\sigma_{\beta_1\beta_2}\cdots\Gamma_{\mu_1}\cdots
\Gamma_{\mu_s}\cdots\sigma_{\beta_{2t-1}\beta_{2t}}$,
where the number of $\sigma$'s
and $\Gamma$'s is the same as in equation (\ref{Sla16}) and the indices
$\beta_1..\mu_1..\mu_s..\beta_{2t}$ form some permutation
of the indices $b_1..b_{2t},m_1..m_s$.

A similar representation holds for the sums
\be
\sum_{ij}\sigma_{ij}\sigma_{b_1b_2}\cdots\Gamma_{m_1}\cdots
\sigma_{b_lb_{l+1}}\cdots\Gamma_{m_s}\cdots
\sigma_{b_{2t-1}b_{2t}}\sigma_{ij},
\label{Sla17}
\ee
\be
\sum_i\sigma_{b_0i}\sigma_{b_1b_2}\cdots\Gamma_{m_1}\cdots
\sigma_{b_lb_{l+1}}\cdots\Gamma_{m_s}\cdots
\sigma_{b_{2t-1}b_{2t}}\sigma_{ib_{2t+1}}.
\label{Sla18}
\ee
Indeed any $\sigma_{ij}$ is a commutator of two $\Gamma$
matrices and applying successively the equation (\ref{simp0})
one gets equations (\ref{Sla17},\ref{Sla18}).
These equations are sufficient to prove our statement for
arbitrary fermion loop, where all the vertices are placed on the
loop, and there are no vertices inside the loop (see figure~2).

Indeed, contraction of Yang-Mills or Higgs fields
produces the terms proportional to the sums
$$
\sum_{ij}
\sigma_{ij}
\sigma_{b_1b_2}\cdots\Gamma_{m_1}\cdots\Gamma_{m_s}\cdots
\sigma_{b_{2t-1}b_{2t}}
\sigma_{ij}.
$$
According to discussion presented above after summation over
''dummy'' indices one gets the sum of products
$$
\sigma_{\beta_1\beta_2}\cdots\Gamma_{\mu_1}\cdots\Gamma_{\mu_s}\cdots
\sigma_{\beta_{2t-1}\beta_{2t}},
$$
where $\beta_1..\beta_{2t}$, $\mu_1..\mu_s$
are some permutation of external fields indices
$b_1..b_{2t+1}$, $m_1..m_s$.

So finally we have to take the traces of the type
\be
{\tr}\Bigl(\frac{1+\Gamma_{11}}{2}
\sigma_{\beta_1\beta_2}\cdots\Gamma_{\mu_1}
\cdots\Gamma_{\mu_s}\cdots\sigma_{\beta_{2t-1}\beta_{2t}}\Bigr),
\label{Sla20}
\ee
where the indices $\beta_1..\beta_{2t},\mu_1..\mu_s$ form
some permutation of external fields indices.
By the same arguments which have been given above for one loop
diagrams, the trace involving $\Gamma_{11}$ is zero if the number
of external vector fields plus one half of the number of external
Higgs fields is less then 5.

Now we extend our proof to fermion loops
with arbitrary boson subdiagram. These diagrams may include
internal vertices describing the selfinteraction of Yang-Mills
fields, selfinteraction of Higgs fields
and their mutual interactions.

The threelinear Yang-Mills interaction looks as follows:
$$
V_{AAA}=i{\sl g}t^{(ij)(kl)(mn)}
[(p-k)_\rho g_{\mu\nu}+(k-q)_\mu g_{\nu\rho}+(q-p)_\nu g_{\mu\rho}],
$$
and the structure constants $t^{(ij)(kl)(mn)}$ are linear
combinations of $\delta$'s:
$$
[\sigma_{ij},\sigma_{kl}]=t^{(mn)(ij)(kl)}\sigma_{mn},
$$
$$
t^{(mn)(ij)(kl)}=i\bigl\{
(im)(ln)(jk)-(km)(jn)(il)-(jm)(nl)(ik)+(km)(in)(lj)-
$$
$$
-(in)(lm)(jk)+(km)(jm)(il)+(jn)(ml)(ik)-(kn)(im)(lj)\bigr\},
$$
where $(ij)=\delta_{ij}$.

The fourlinear interaction has a form:
$$
V_{AAAA}={\sl g}^2\{t^{(ab)(cd)(ij)}t^{(ij)(ef)(hk)}
(g_{\mu\rho}g_{\nu\sigma}-g_{\mu\sigma}g_{\nu\rho})+permutations\}.
$$
Again the SO(10) structure constants $t^{(ab)(cd)(ij)}t^{(ij)(ef)(hk)}$
consist of $\delta$'s: $\delta_{bc}\delta_{de}\delta_{fh}\delta_{ka}~+~\ldots$

The vertices including the Higgs fields possess analogous property --
their SO(10) structure is given by a linear combinations
of $\delta$-functions:
$$
V_{AA\phi\phi}={\sl g}^2\{
  \delta^{be} \delta^{df} \delta^{ac}
- \delta^{ae} \delta^{df} \delta^{bc}
- \delta^{be} \delta^{cf} \delta^{ad}
+ \delta^{ae} \delta^{cf} \delta^{bd}\},
$$
$$
V_{\phi\phi\phi\phi}=-8\lambda\{
\delta^{ef}\delta^{hj}+
\delta^{eh}\delta^{fj}+
\delta^{ej}\delta^{fh}\},
$$
$$
V_{A\phi\phi}=i{\sl g}\{
k_\mu\delta^{ea}\delta^{fb}-p_\mu\delta^{eb}\delta^{fa}\},
$$
where $A$'s carry SO(10)-indices $a,b$ and $c,d;$ $\phi$'s
have indices $e,f,h$ and $j$.

Therefore considering an arbitrary diagram without internal
spinor lines we may firstly perform the summation over
all ''dummy'' indices. After this summation we shall
get the sum of the products of the type (\ref{Sla20}),
where as above the indices $\beta_1..\beta_{2t},\mu_1..\mu_s$
form some permutation of external lines indices. Hence
we arrive to the same conclusion: only diagrams with
the number of external vector lines plus
one half of the external Higgs lines less the 5 may depend
on $\Gamma_{11}.$

The theorem does not hold for the diagrams with one fermion loop
and open fermion lines (see for example figures 3 and 4).

\medskip
\Lengthunit=1.0cm
\Linewidth{0.5pt}
\GRAPH(hsize=9){%
\Linewidth{0.5pt}%
\mov(0.2,2){\wavelin(0.8,0)}%
\mov(0.2,2.4){\wavelin(0.87,0)}%
\mov(0.2,1.6){\wavelin(0.87,0)}%
\mov(2.93,2.4){\wavelin(0.67,0)}%
\mov(2.93,1.6){\wavelin(0.67,0)}%
\mov(2,2.0){\Ellipse*(1)[1.4,1]}%
\mov(1.35,1.85){Bosons}%
\mov(1.3,1.3){\wavelin1.04(0.5,0.2)}%
\mov(2.7,1.3){\dashlin1.04(-0.5,0.2)}%
\mov(1.3,2.7){\dashlin1.04(0.5,-0.2)}%
\mov(2.7,2.7){\wavelin1.04(-0.5,-0.2)}%
\Linewidth{0.8pt}%
\mov(2,2.0){\Circle(2)}%
\mov(3.6,1.6){\lin(0.0,0.8)}%
\mov(3.6,1.6){\lin(0.4,-0.6)}%
\mov(3.6,2.4){\lin(0.4,0.6)}%
\mov(1.5,0.2){\normalsize Figure 3.}%
\Linewidth{0.5pt}%
\mov(4.7,2){\wavelin(0.8,0)}%
\mov(7.32,2.6){\wavelin(0.78,0)}%
\mov(7.49,2.2){\wavelin(0.61,0)}%
\mov(7.49,1.8){\wavelin(0.61,0)}%
\mov(7.32,1.4){\wavelin(0.78,0)}%
\mov(6.5,2.0){\Ellipse*(1)[1.4,1]}%
\mov(5.85,1.85){Bosons}%
\mov(5.8,1.3){\dashlin1.04(0.5,0.2)}%
\mov(7.2,1.3){\wavelin1.04(-0.5,0.2)}%
\mov(5.8,2.7){\wavelin1.04(0.5,-0.2)}%
\mov(7.2,2.7){\dashlin1.04(-0.5,-0.2)}%
\Linewidth{0.8pt}%
\mov(6.5,2.0){\Circle(2)}%
\mov(8.1,1.4){\lin(0.0,1.2)}%
\mov(8.1,1.4){\lin(0.3,-0.4)}%
\mov(8.1,2.6){\lin(0.3,0.4)}%
\mov(6.0,0.2){\normalsize Figure 4.}%
\Linewidth{0.5pt}%
\mov(9.2,2.3){\wavelin(0.75,0)}%
\mov(9.2,1.7){\wavelin(0.75,0)}%
\mov(12.45,2.3){\wavelin(0.75,0)}%
\mov(12.45,1.7){\wavelin(0.75,0)}%
\mov(10.75,2.5){\wavelin(0.88,0)}%
\mov(10.9,2){\wavelin(0.6,0)}%
\mov(10.75,1.5){\wavelin(0.88,0)}%
\Linewidth{0.8pt}%
\mov(10.4,2.0){\Ellipse(1)[1,1.4]}%
\mov(12,2.0){\Ellipse(1)[1,1.4]}%
\mov(10.5,0.2){\normalsize Figure 5.}%
}
\normalsize

However the diagrams shown at figure 3 are superficially
convergent and do not include divergent fermion loops.
Hence they may be treated as was explained above.
One firstly applies dimensional regularization to calculate the
counterterms corresponding to divergent subgraphs,
then makes a subtraction according to R-operation.
After that the diagram becomes convergent and allows
continuation to $d=4$.
The calculation of the trace over the fermion loop may be done in
four dimensions where no problems arise.

The only diagrams of this type which cannot be treated in this way
are the diagrams with one external vector line and two external fermion
lines including a fermion loop with more then four bosonic lines
(see for example figure 4).
These diagrams are superficially divergent and one cannot remove
dimensional regularization before calculating all the traces.
The lowest order diagram of this type contains four loops.

The diagrams including more then one fermion loop, like the
diagram shown at figure 5 also cannot be treated with the
dimensional regularization described above. For their
gauge invariant analysis one has to apply the hybrid
regularization described in the section III. The lowest
order diagram of this type also contains four loops.

\end{document}